\documentclass{moriond}

\def\al{\alpha}

\def\be{\begin{equation}}
\def\ee{\end{equation}}
\def\bea{\begin{eqnarray}}
\def\eea{\end{eqnarray}}

\newcommand{\mR}{\mathcal{R}}

\newcommand{\divA}{\nabla \cdot A}

\newcommand{\hGam}{\hat{ \Gamma}}

\newcommand{\Fd}{\tilde{F}}

\newcommand{\bt}{\beta}
\newcommand{\ga}{\gamma}

\newcommand{\lp}{\left(}
\newcommand{\rp}{\right)}

\newcommand{\Lag}{{\mathcal L}}

\newcommand{\lsim}   {\mathrel{\mathop{\kern 0pt \rlap
  {\raise.2ex\hbox{$<$}}}
  \lower.9ex\hbox{\kern-.190em $\sim$}}}
\newcommand{\gsim}   {\mathrel{\mathop{\kern 0pt \rlap
  {\raise.2ex\hbox{$>$}}}
  \lower.9ex\hbox{\kern-.190em $\sim$}}}

\newcommand{\Photo}{}

\begin{document}
\vspace*{4cm}
\title{Cosmology with vector distortion}

\author{ Jose Beltr\'an Jim\'enez }

\address{Aix Marseille Univ, Univ Toulon, CNRS, CPT, Marseille, France.}

\maketitle\abstracts{We consider an extension of Weyl geometry with the most general connection linearly determined by a vector field. We discuss some of the geometrical properties within this framework and then we construct gravitational theories leading to an interesting class of vector-tensor theories with cosmological applications.
}

\section{Introduction}
The beauty of General Relativity (GR) resides in its geometrical interpretation as the curvature of the spacetime. A foundational property of GR is that the metric determines the entirely metric and affine structures of the spacetime. However,  these two structures do not need to be related and, thus, if we decide to embrace the geometrical description of gravity, we are naturally led to considering the potential role of the neglected dof's contained in the connection. For the Einstein-Hilbert action, the connection is {\it dynamically} fixed to be metric compatible, but for more general gravitational theories, the inclusion of an arbitrary connection does lead to different physical effects \cite{Olmo:2011uz}. These additional dof's are associated to the presence of torsion  and/or non-metricity in the spacetime \cite{Hammond:2002rm}. An intermediate approach is to consider only a constrained sector of the connection. A paradigmatic example of this is the Weyl geometry \cite{Scholz:2011za}, where only the (vector) trace of the non-metricity is allowed. Aspects of Weyl geometry have been studied \cite{Tanhayi:2012nn,Jimenez:2014rna} and it represents a natural arena to formulate conformally invariant theories. We will introduce an extended version of this geometry and gravitational theories within this framework.

\section{Extending Weyl geometry}
Soon after Einstein formulated GR, Weyl considered an extension of its geometrical framework where the metric-compatibility condition was replaced by $\hat{\nabla}_{\mu} g_{\alpha\beta} =-2 A_\mu g_{\alpha\beta}$ with $A_\mu$ some vector field. This condition is invariant under the Weyl rescaling $g_{\mu\nu}\rightarrow e^{2\Lambda(x)}g_{\mu\nu}$ together with $A_\mu\rightarrow A_\mu-\partial_\mu\Lambda(x)$ and we can easily solve for the connection as (assuming vanishing torsion)
\be
\hGam^\alpha_{\beta\gamma}=\Gamma^\alpha_{\beta\gamma}-\left(A^\alpha g_{\beta\gamma}-2A_{(\beta}\delta^\alpha_{\gamma)}\right)\,,
\label{Wconnection}
\ee
with $\Gamma^\alpha_{\beta\gamma}$ the usual Levi-Civita connection. This connection inherits the aforementioned Weyl symmetry and, therefore, the associated curvature tensor will also be invariant under such transformation. This Weyl connection can be naturally extended to the most general affine structure {\it linearly} determined by one single vector field and with no derivatives, which can be expressed as\cite{Jimenez:2015fva}
\be \label{vgamma}
\hat{\Gamma}^\al_{\bt\ga} = \Gamma^\al_{\bt\ga}  
-b_1A^\alpha g_{\beta\gamma}+b_2\delta^\alpha_{(\beta} A_{\gamma)}+b_3\delta^\alpha_{[\beta} A_{\gamma]}\,
\ee
with $b_i$  some parameters. The non-Levi-Civita part of the connection is called {\it distortion} and, in this case, it carries both non-metricity and torsion. The metric (in-)compatibility condition in this case reads
\be
\hat{\nabla}_\mu g_{\alpha\beta}=(b_3-b_2)A_\mu g_{\alpha\beta} + (2b_1-b_2-b_3) A_{(\alpha}g_{\beta)\mu}\,.
\label{nonmetricity2}
\ee
For geometries with $2b_1-b_2-b_3=0$, this expression also remains invariant under a Weyl transformation so that they represent a generalization of Weyl geometry with a torsion component determined by the vector field. As we will discuss below, these geometries turn out to be special when constructing gravitational Lagrangians.

\section{Gravitational theories}
The curvature of the spacetime within the framework of geometries with vector distortion depends on the vector field and, thus, actions expressed in terms of curvature invariants will generically lead to theories with non-minimally coupled vector fields. Here we will consider two particularly interesting examples.

\subsection{$f(\mR)$ theories}
The Ricci scalar for the connection with vector distortion reads $
\mR=R-\bt_1 A^2+\beta_2\divA\,$,
with $
\beta_1  \equiv-3[4b_1^2-8b_1(b_2+b_3)+(b_2+b_3)^2]/4$ and $
\beta_2 \equiv-3(2b_1+b_2+b_3)/2$. If we take a Lagrangian to be an arbitrary function of this Ricci scalar $\Lag=f(\mR)$, one can show that the vector is forced to be the gradient of a scalar so that only one dof will propagate. The Lagrangian for the resulting scalar can be written as \cite{Jimenez:2015fva}
\be \label{nbd}
\mathcal{L} = \varphi R + \frac{\beta_2^2}{4\beta_1\varphi}\partial_\mu\varphi\partial^\mu\varphi -V(\varphi)\,
\ee
where $V(\varphi)$ is determined by the functional form of $f$. Here we can recognize the form of a Brans-Dicke scalar-tensor theory. In the special case of the generalized Weyl connection with $b_3=2b_1-b_2$ we find  $\beta_2^2=6\beta_1$ which corresponds to a non-propagating scalar field (resembling the usual results for $f(\mR)$ theories in the Palatini formalism). In the particular case of a quadratic function, the resulting Lagrangian can be written in the Einstein frame as
\be \label{efr}
\Lag=  \frac{M_{pl}^2}{2}\tilde{R}-\frac{1}{2}\tilde{g}^{\mu\nu}\tilde{\phi}_{,\mu}\tilde{\phi}_{,\nu} - \frac{3}{4}M_{pl}^2 M^2\lp 1 - e^{-\sqrt{\frac{2}{3\al}}\frac{\tilde{\phi}}{M_{pl}}}\rp^2,
\ee
with $M^2$ the new mass scale and $\alpha\equiv 1-\beta_2^2/(6\beta_1)$. The potential for the scalar field defines a one-parameter family of potentials that generalize the Starobinsky inflationary model (which is recovered for $\alpha\rightarrow 1$). For arbitrary $\alpha$, we obtain the so-called $\alpha$-attractor parameterization that appears in classes of supergravity and superconformal inspired inflationary models \cite{Kallosh:2013yoa}. This vector field formulation for the $\alpha$-attractor model has also been considered within the context of modified gravity with auxiliary vector fields \cite{Ozkan:2015iva}, but without the geometrical framework of connections with vector distortion.

\subsection{More general Lagrangians}

In the previous subsection we have discussed how general Lagrangians constructed out of the Ricci scalar lead to scalar-tensor theories. In order to obtain proper vector-tensor theories (which do not reduce to one scalar dof), it is necessary to include more general curvature invariants. Motivated by Gauss-Bonnet terms we can seek for terms quadratic in the curvature. We can thus consider the most general quadratic Lagrangian in the Riemann tensor. Since we want to recover a healthy theory for vanishing distortion, we will impose that it reduces to pure Gauss-Bonnet when $A_\mu=0$. Once everything is expressed in terms of the Levi-Civita connection and the vector field, the Lagrangian can be written as \cite{Jimenez:2015fva,Jimenez:2016opp}
\be
\Lag_{(2)}=-\frac1 4 F_{\mu\nu} F^{\mu\nu}+\xi A^2\divA -\lambda A^4-\beta G^{\mu\nu} A_\mu A_\nu+
\gamma_1(\divA)^2+\big(\gamma_2 A^2+\gamma_3\divA\big)R
\ee
with $F_{\mu\nu}=\partial_\mu A_\nu-\partial_\nu A_\mu$ and $\xi$, $\lambda$, $\beta$ and $\gamma_i$ some parameters that depend on the distortion parameters and the coefficients of the quadratic terms in the Lagrangian. The above Lagrangian propagates 4 dof's for the vector field, one of which will be affected by the Ostrogradski instability. In order to avoid it, we impose the conditions $\gamma_i=0$, which (in dimensions higher than 3) uniquely determine the distortion coefficients to satisfy \cite{Jimenez:2016opp} $b_3=2b_1-b_2$, i.e., only the generalized Weyl class will give rise to healthy theories. If we add the Einstein-Hilbert term for the distorted connection we will also generate a mass term for the vector so we finally find
\be
\Lag_{(2)}=-\frac1 4 F_{\mu\nu} F^{\mu\nu}+\frac12 M^2 A^2-\lambda A^4+\xi A^2\divA -\beta G^{\mu\nu} A_\mu A_\nu
\label{VT}
\ee
which was already obtained within the context of pure Weyl geometry \cite{Jimenez:2014rna}. The first two terms simply represent the Proca lagrangian for a massive vector field and (for $\xi=\lambda=\beta=0$) it was suggested \cite{Jimenez:2014rna} that they could be a natural vector dark matter candidate. Since the mass can be close to the Planck mass, the field will be very heavy and, thus, we can apply the isotropy theorem \cite{Cembranos:2012kk} so that the corresponding averaged energy-momentum tensor will be isotropic and will mimic a dust component. The $\xi$-term is a vector Galileon interaction, which is a healthy vector non-gauge invariant derivative self-interaction, and the $\beta$-term is a healthy non-minimal coupling for the vector field, both particular examples of more general classes \cite{Tasinato:2014eka}. The coupling to the Einstein tensor was explored as a potential mechanism to generate cosmic magnetic fields \cite{BeltranJimenez:2010uh}. The vector-tensor theory described by (\ref{VT}) has been shown to have interesting cosmological applications \cite{Jimenez:2016opp}. It has isotropic de Sitter phases, bouncing/re-collapsing solutions, the possibility of self-tuning or singularities with divergent $\dot{H}$, but finite $H$ and $\rho$. It was also shown that the coupling to the Einstein tensor generates an anomalous propagation speed for gravitational waves and, therefore, it is subject to the corresponding constraints from binary pulsars \cite{Jimenez:2015bwa}.

At cubic order, we can use the result by Horndeski \cite{Horndeski:1976gi} that the non-minimal coupling $L^{\mu\nu\alpha\beta} F_{\mu\nu} F_{\alpha\beta}$, with $L^{\mu\nu\alpha\beta}=-\frac12 \epsilon^{\ \mu\nu\rho\sigma}\epsilon^{\alpha\beta\gamma\delta} R_{\rho\sigma\gamma\delta}$ the double dual Riemann tensor leads to second order equations of motion. Thus, we can use this term for our connection with vector distortion to generate additional interactions, which can then be written as \cite{Jimenez:2015fva,Jimenez:2016opp}
\bea
\Lag_{(3)}&=&L^{\mu\nu\alpha\beta} F_{\mu\nu} F_{\alpha\beta}+2(2b_1+b_2+b_3)\Fd^{\mu\alpha}\Fd^\nu{}_\alpha\nabla_\mu A_\nu\nonumber\\
&&+\frac{1}{2}\Big[\Big(2b_1-b_2-b_3\Big)^2A^2g^{\mu\nu}-2\Big(4b_1^2+(b_2+b_3)^2\Big)A^\mu A^\nu\Big]F_{\mu\alpha} F_\nu{}^\alpha\;.
\eea
In the first line we find the vector-tensor Horndeski interaction\cite{Horndeski:1976gi} and a healthy (non-gauge invariant) derivative self-interaction, while in the second line we have standard interactions for the vector field. The cosmology and stability of the Horndeski interaction has been studied \cite{Barrow:2012ay}, showing that its relevant effects typically come in associated with instabilities.

\section{Conclusions}
Geometries with vector distortion represent a suitable framework to formulate gravitational theories leading to interesting classes of vector-tensor theories. They can easily accommodate the $\alpha$-attractor generalization of Starobinsky inflation with the vector field effectively describing a scalar field. More general actions lead to ghost-free non-gauge invariant derivative self-interactions for the vector which can support de Sitter solutions and, thus, be used as candidates for inflation and/or dark energy. The presence of bouncing/re-collapsing solutions also give a novel scenario to consider bouncing  and/or ekpyrotic universes.

\section*{Acknowledgments}
It is a pleasure to thank the organizers of the Rencontres de Moriond 2016 and their participants for a memorable week. I also
acknowledge the financial support
of A*MIDEX project (n ANR-11-IDEX-0001-02) funded by the ÓInvestissements dÕAvenirÓ French Government program,
managed by the French National Research Agency (ANR), MINECO (Spain) projects FIS2014-52837-P and
Consolider-Ingenio MULTIDARK CSD2009-00064%

\end{document}